\documentclass[letter]{jpsj2} 
\usepackage{latexsym}
\usepackage{bm}
\usepackage{graphicx}

\def\slashchar#1{\setbox0=\hbox{$#1$} 
\dimen0=\wd0 
\setbox1=\hbox{/} \dimen1=\wd1 
\ifdim\dimen0>\dimen1 
\rlap{\hbox to \dimen0{\hfil/\hfil}} 
#1 
\else 
\rlap{\hbox to \dimen1{\hfil$#1$\hfil}} 
/ 
\fi}

\renewcommand{\slashchar}{}

\title{
Topological Stability of Majorana Zero Modes in 
Superconductor-Topological Insulator Systems
}

\author{Takahiro Fukui and Takanori Fujiwara}
\inst{
Department of Physics, Ibaraki University, Mito
310-8512 
}

\recdate{\today}

\abst{
We derive an index theorem for 
zero-energy Majorana fermion modes in a superconductor-topological insulator system
in both two and three dimensions, which is valid for
models with chiral symmetry as well as particle-hole symmetry.
For more generic models without chiral symmetry, we suggest that 
Majorana zero modes are classified by Z$_2$.
}

\kword{
Majorana zero modes, superconductor-topological insulator system,
index theorem, topological invariants
}

\begin{document}
\maketitle

Majorana bound states have recently been attracting much interest as elementary
excitations obeying non-Abelian statistics \cite{ReaGre00}.
They have been predicted in $\nu=5/2$ quantum Hall effects, \cite{ReaGre00}
$p$-wave superconductors, \cite{ReaGre00,Ivanov01,SOM04,SNT06,TSL07}
topological insulators with the proximity effect of $s$-wave 
superconductors\cite{FuKan08} as well as of ferromagnetic insulators,\cite{TYN09} 
$s$-wave superfluids of ultracold fermionic atoms\cite{STF09}, 
vacuum states of superfluid $^3$He-B,\cite{Vol09}
superconducting states in graphene, \cite{Her09-1} 
and massive Dirac equations in magnetic fields.\cite{Her09-2}
Recently, Teo and Kane \cite{TeoKan09} have proposed a model of 
superconductor-topological insulator systems in three dimensions. 
They have shown that the third Chern 
number, which is a topological invariant for three-dimensional topological 
insulators, \cite{QHZ08} reduces to a winding number of order parameters on 
a two-dimensional sphere surrounding a pointlike topological defect.
This suggests that the Majorana zero modes have
a topological origin and are stably protected from perturbations.
This property may play a crucial role in the application of Majorana zero modes
to topological quantum computations.\cite{Kitaev03,NSSFS08}
Along this line, Tewari {\it et. al.} have established a theorem for the 
existence of Majorana zero modes in chiral $p$-wave superconductors\cite{TSL07} 
and in spin-orbit-coupled semiconductors.\cite{TSS09}

In this paper, we derive an index theorem for the Majorana zero modes in
a model of superconductor-topological insulator systems.
To be concrete,  
we establish the relationship between the sum of chiralities of the Majorana 
zero modes and a topological invariant associated with order parameters.
This indicates that the Majorana zero modes in such 
systems are stable owing to their topological origin. 
When we derive the above index theorem, we use a simplified model that has 
not only particle-hole symmetry, as the Bogoliubov-de Gennes Hamiltonians 
should have, but also an additional chiral symmetry.
We next discuss the effects of chiral symmetry breaking perturbations,
especially those of the chemical potential term.
We suggest that the number of zero modes is classified by Z$_2$. 
Namely, only the evenness or oddness of the number of zero modes
in models with chiral symmetry is invariant 
under chiral symmetry breaking perturbations.
This is consistent with the results obtained by Tewari {\it et. al.}\cite{TSL07}
Finally, we present a concrete zero mode wave function for the Fu-Kane
model \cite{FuKan08} in both the cases of $\mu=0$ and $\mu\ne0$.

Let ${\cal H}$ be a Hamiltonian defined by
\begin{alignat}1
{\cal H}=i\slashchar{{\cal D}}=i\gamma^j\partial_j+\gamma^a\phi_a,
\label{Ham}
\end{alignat}
where $j=1,\cdots,d$ and $a=d+1,\cdots,2d$, and
$\gamma^\mu$ ($\mu=1,\cdots,2d$) are $\gamma$ matrices in $2d$ dimensions 
$\{\gamma^\mu,\gamma^\nu\}=2\delta^{\mu\nu}$.
$\bm\phi$ stands for a set of order parameters.
This Hamiltonian has been 
introduced by Kane and his collaborators \cite{FuKan08,TeoKan09}
as a minimal model 
describing superconductor-topological insulator systems.
In the case of $d=2$, $\bm\phi=\Delta_0(r)(\cos \theta,\sin \theta)$
is the model proposed by Fu and Kane \cite{FuKan08} 
to study a Majorana zero mode in a vortex of an $s$-wave superconductor,
where $(r,\theta)$ are the polar coordinates in two dimensions.
In the case of $d=3$, $\bm\phi=({\rm Re}\,\Delta,{\rm Im}\,\Delta,m)$ 
describes generic superconductor-topological insulator systems studied by 
Teo and Kane \cite{TeoKan09}, where the region $m<0$ corresponds to a
bulk topological insulator, with the proximity effect $\Delta\ne0$
induced by a superconductor. 
If $\bm\phi\ne0$ is constant, the spectrum of the Hamiltonian has a gap,
describing a massive Dirac particle. Even with a more generic $\bm\phi$ 
that is not 
spatially uniform,
the spectrum may still be massive. However, if $\bm\phi$ has
a topological defect, there appear bound states around it.
In particular, it has been shown\cite{FuKan08,TeoKan09} 
that the present model (\ref{Ham}) can admit generically 
bound states sitting exactly at zero energy. 
To study such zero modes bound to a pointlike defect, 
we assume that away from the defect located at $x=0$, 
the asymptotic form of the order parameter $\bm\phi$ is given by 
\begin{alignat}1
|\bm\phi|\rightarrow \phi_0=\mbox{const. }\quad (|x|\rightarrow \infty).
\label{AsyPhi}
\end{alignat}
The explicit expression for $\gamma$ matrices is not 
necessary,\cite{footnote1} 
but we have to arrange them to obey
$C\gamma^jC^{-1}=\gamma^j$ and 
$C\gamma^aC^{-1}=-\gamma^a$,
where $C$ (${C}^2=1$) is the {\it anti}-unitary operator \cite{footnote1} 
describing particle-hole symmetry.
Therefore, the Hamiltonian (\ref{Ham}) has particle-hole symmetry, 
\begin{alignat}1
C{\cal H}C^{-1}=-{\cal H},
\label{ParHol}
\end{alignat}
as it should be in the case of the Bogoliubov-de Gennes Hamiltonians.
The wave function of the zero modes can be chosen such that
$\varphi_{0,i}(x)=C\varphi_{0,j}(x)$, 
which has been referred to as a Majorana fermion mode.
Another symmetry of the Hamiltonian (\ref{Ham}) is chiral symmetry denoted by
\begin{alignat}1
\gamma_5{\cal H}\gamma_5=-{\cal H} ,
\label{Chi}
\end{alignat}
where $\gamma_5=(-i)^d\gamma^1\cdots\gamma^{2d}$ with the property
$C\gamma_5C^{-1}=\gamma_5$.
It should be noted that the $\gamma_5$ matrix is that in $2d$ 
dimensions in the ordinary sense, 
although we are dealing with a model in $d$ spatial dimensions.
This symmetry seems artificial, since the chemical potential term 
and many other terms with some products of $\gamma$ matrices 
break chiral symmetry. 
According to the Altland-Zirnbauer classification scheme \cite{Zir96,AltZir97},
the model only with particle-hole symmetry (\ref{ParHol}) is classified as 
class D, which is the most generic class of superconductors, 
whereas the model with an additional symmetry (\ref{Chi}) belongs to class BDI,
which is the time reversal invariant chiral class.\cite{SRFL08,Kitaev08}

Even with such an enhanced symmetry (and although we have neglected background gauge fields
for simplicity), the minimal model (\ref{Ham}) is nevertheless 
useful for clarifying the topological stability of the Majorana zero modes,
as we shall show momentarily.
When $d=2$, it has a deep relationship with a
Dirac fermion coupled to a Higgs field with a vortex background \cite{JacRos81}.
Jackiw and Rossi \cite{JacRos81} have derived the zero mode 
wave functions of this model explicitly. 
It has been shown by Weinberg \cite{Wei81}
that these zero modes are indeed 
ensured by the index theorem, even though this model is defined for
the noncompact two-dimensional space R$^2$.
The model in three dimensions is associated with a Dirac fermion coupled to a Higgs field with 
a 't Hooft-Polyakov monopole \cite{tHooft,Polyakov} 
background whose index theorem, surprisingly valid for open and odd-dimensional spaces,
has been derived by Callias \cite{Callias78}.
In what follows, we first prove the index theorem for the Hamiltonian (\ref{Ham})
simultaneously in $d=2$ and $d=3$ cases using the method developed by Weinberg\cite{Wei81},
and next, we discuss the effect of the chemical potential 
that breaks chiral symmetry.

Let us first define the index of the Hamiltonian (\ref{Ham}) by
${\rm ind}\,{\cal H}=N_+-N_-$, where $N_\pm$ denote, respectively, the number of 
zero modes of (\ref{Ham}), ${\cal H}\varphi_{0,i}^\pm=0$, with $\pm$ chirality,
$\gamma_5\varphi_{0,i}^\pm=\pm\varphi_{0,i}^\pm$.
This can be written as
\begin{alignat}1
{\rm ind}\,{\cal H}  
=\lim_{m\rightarrow0}{\rm Tr}\,\gamma_5\frac{m^2}{-\slashchar{{\cal D}}^2+m^2} ,
\label{Ind}
\end{alignat}
where Tr denotes the trace over the space coordinates as well as the $\gamma$
matrices. 
On the other hand, it is well-known that in the large mass limit, the same 
quantity yields the topological invariant
\begin{alignat}1
c_d=\lim_{M\rightarrow\infty}
{\rm Tr}\,\gamma_5\frac{M^2}{-\slashchar{{\cal D}}^2+M^2} ,
\label{TopInv}
\end{alignat}
where $c_d$ is the $d$th Chern number \cite{FujSuz05} if $d=$ even, while
$c_d=0$ if $d=$ odd.
It is possible to express Tr in both eqs. (\ref{Ind}) and (\ref{TopInv})
using the eigenstates of the Hamiltonian,
${\cal H}\varphi_{n}=\lambda_n\varphi_n$. In the case of nonzero energies, 
$\varphi_{-n}\equiv\gamma_5\varphi_n$ has the energy $\lambda_{-n}\equiv-\lambda_n$,
and hence, these two states are orthogonal to each other, 
$\int d^dx\varphi_n^\dagger\gamma_5\varphi_n=\int d^dx\varphi_n^\dagger\varphi_{-n}=0$.
Then, it is easy to see that eqs. (\ref{Ind}) and (\ref{TopInv}) give the same 
index of the Hamiltonian. However, in the plane-wave basis, the above two 
quantities become two kinds of different invariants, giving
a nontrivial relationship between them.
The necessity of two kinds of masses is manifested  
when the axial-vector current is introduced such that
\begin{alignat}1
J^i(x,m,M)&=
\lim_{y\rightarrow x}{\rm tr}\,\gamma_5\gamma^i
\left(
\frac{1}{\slashchar{{\cal D}}+m}-
\frac{1}{\slashchar{{\cal D}}+M}
\right)
\delta(x-y) 
\nonumber\\
&=-\lim_{y\rightarrow x}{\rm tr}\,\gamma_5\gamma^i\slashchar{{\cal D}}
\left(
\frac{1}{-\slashchar{{\cal D}}^2+m^2}-
\frac{1}{-\slashchar{{\cal D}}^2+M^2}
\right)
\delta(x-y)  ,
\label{AxiCur}
\end{alignat}
where tr denotes the trace over the $\gamma$ matrices, and in deriving
the second line, we have used the fact that $\slashchar{{\cal D}}$ 
anticommutes with $\gamma_5$.
The first term is, in the limit of $m\rightarrow0$, the axial-vector current of the 
model under consideration, while 
the second term is the Pauli-Villars regulator whose mass $M$
is set to infinity after the calculation. Such a regulator field is
needed to make the current well-defined.

It is well-known that the index (\ref{Ind})
and the Chern number (\ref{TopInv}) have an intimate relationship 
with the chiral anomaly, which yields an anomalous term to the conservation law of 
the axial-vector current. 
To see this, it is convenient to describe the current on the basis of 
eigenstates of the Hamiltonian. The completeness of eigenstates
of the hermitian Hamiltonian,
$\sum_n\varphi_n(x)\varphi_n^\dagger(y)=\delta(x-y)$, leads to
\begin{alignat}1
J^i(x,m,M)
&=-\sum_n\varphi_n^\dagger(x)\gamma_5\gamma^i(-i\lambda_n)
\left(
\frac{1}{\lambda_n^2+m^2}-
\frac{1}{\lambda_n^2+M^2}
\right)\varphi_n(x) .
\nonumber
\end{alignat}
Then, the divergence of the current is easy to compute, since
the derivative $\partial_i$ operates only on the wave functions 
in the above expression. After some calculations, in particular, using the
relation $\gamma^j\partial_j\varphi_n=i(\gamma^a\phi_a-\lambda_n)\varphi_n$,
we obtain
\begin{alignat}1
\partial_iJ^i(x,m,M)
&=-2\sum_n\varphi_n^\dagger(x)\gamma_5
\left(
\frac{m^2}{\lambda_n^2+m^2}-
\frac{M^2}{\lambda_n^2+M^2}
\right)\varphi_n(x) .
\nonumber
\end{alignat}
It thus turns out that the divergence of the current (\ref{AxiCur}) is given by 
\begin{alignat}1
\partial_iJ^i(x,m,M)=\lim_{y\rightarrow x}(-2){\rm tr}\,\gamma_5
\left(
\frac{m^2}{-\slashchar{{\cal D}}^2+m^2}-
\frac{M^2}{-\slashchar{{\cal D}}^2+M^2} 
\right)
\delta(x-y) .
\nonumber
\end{alignat}
Integrating both sides over the space R$^d$ and 
taking the limits $m\rightarrow0$ and $M\rightarrow\infty$ yield
\begin{alignat}1
{\rm ind}\,{\cal H}-c_d&=-\frac{1}{2}\int d^dx\partial_iJ^i(x,0,\infty)
\nonumber\\
&=-\frac{1}{2}\int dS_iJ^i(x,0,\infty),
\label{ModIndThe}
\end{alignat}
where $dS_i$ is the infinitesimal surface element on S$^{d-1}$, i.e., the boundary 
of R$^d$ at $|x|\rightarrow\infty$.
This equation indicates that in even dimensions if the space manifold is closed, 
the surface term above vanishes, and 
the index of ${\cal H}$ is just given by the Chern number $c_d$.
For generic open manifolds, the index theorem is modified by the surface term
given directly by the axial-vector current.
It should be noted that in the present model (\ref{Ham}), $c_d$ always vanishes since
we have neglected the gauge fields, for simplicity, and 
thus, the index of the Hamiltonian, which is due to the Majorana zero modes, 
is given not by the Chern number, but by the surface term in eq. (\ref{ModIndThe}). 
This surface term is also a topological number associated with the configuration of 
the order parameter $\bm\phi$ at $|x|\rightarrow\infty$, as we shall see below.

To compute the surface term, we now calculate the axial-vector current
(\ref{AxiCur}) in the plane-wave basis.
To this end, note 
$
-\slashchar{{\cal D}}^2=-\partial_j^2+\phi_a^2+\gamma^j\gamma^a(i\partial_j\phi_a)
$.
Therefore, at $|x|\rightarrow\infty$, we have
\begin{alignat}1
e^{-ikx}(-\slashchar{{\cal D}}^2)e^{ikx}
&=k_j^2+\phi_0^2+\gamma^j\gamma^a(i\partial_j\phi_a)-\partial_j^2-2ik^j\partial_j,
\nonumber
\end{alignat}
where we have used eq. (\ref{AsyPhi}). This leads to
\begin{alignat}1
J^i(x,m,M)
&=-\int\frac{d^dk}{(2\pi)^d}{\rm tr}\,\gamma_5\gamma^i e^{-ikx}
\slashchar{{\cal D}}\left(\frac{1}{-\slashchar{{\cal D}}^2+m^2}
-\frac{1}{-\slashchar{{\cal D}}^2+M^2}\right)e^{ikx}
\nonumber\\
&=-\int\frac{d^dk}{(2\pi)^d}{\rm tr}\,\gamma_5\gamma^i
\left(\slashchar{{\cal D}}+i\gamma^j\slashchar{k}_j\right)
\nonumber\\
&\times
\left(
\frac{1}{k_j^2+\phi_0^2+m^2+\gamma^j\gamma^a(i\partial_j\phi_a)+{\cal O}}
-\frac{1}{k_j^2+\phi_0^2+M^2+\gamma^j\gamma^a(i\partial_j\phi_a)+{\cal O}}
\right) 
\nonumber\\
&=-\int\frac{d^dk}{(2\pi)^d}{\rm tr}\,\gamma_5\gamma^i
\left(\slashchar{{\cal D}}+i\gamma^j\slashchar{k_j}\right)\sum_{n=0}^\infty(-)^n
\left[
\frac{1}{(k_j^2+\phi_0^2+m^2)^{n+1}}
-\frac{1}{(k_j^2+\phi_0^2+M^2)^{n+1}} \right]
\nonumber\\
&
\times
\left[\gamma^j\gamma^a(i\partial_j\phi_a)+{\cal O}\right]^n ,
\nonumber
\end{alignat}
where ${\cal O}\equiv -\partial_j^2-2ik^j\partial_j$.
Note that ${\rm tr}\,\gamma_5\Gamma$, 
where $\Gamma$ is a product of $\gamma$ matrices 
$\Gamma\equiv\gamma^\mu\gamma^\nu\cdots$,
can be nonzero only when $\Gamma$ includes at least as many as $2d$ $\gamma$ matrices,
which occurs in the $n=1$ ($n=2$) term for $d=2$ ($d=3$) in the 
above expansion. 
Using 
${\rm tr}\,\gamma_5\gamma^{\mu_1}\gamma^{\nu_1}\cdots\gamma^{\mu_d}\gamma^{\nu_d}
=(2i)^d\epsilon^{\mu_1\nu_1\cdots\mu_d\nu_d}$,
we have in the case of $d=2$, 
\begin{alignat}1
J^i(x,0,\infty)
&=\frac{1}{4\pi\phi_0^2}{\rm tr}\,\gamma_5\gamma^i(-i\gamma^a\phi_a)
\gamma^j\gamma^b(i\partial_j\phi_b) +O(|x|^{-2})
\nonumber\\
&=\frac{1}{\pi}\epsilon^{ij}\epsilon^{ab}\hat\phi_a\partial_j\hat\phi_b
+O(|x|^{-2}) ,
\nonumber
\end{alignat}
where $\hat\phi_a=\phi_a/\phi_0$ is the normalized field,
$|\hat{\bm\phi}|=1$, at $|x|\rightarrow\infty$. 
Note here that $\partial_i\phi_a\sim O(|x|^{-1})$ 
since $|{\bm\phi}|$ approaches a constant value.
In the case of $d=3$,
\begin{alignat}1
J^i(x,0,\infty)
&=-\frac{1}{32\pi\phi_0^3}
{\rm tr}\,\gamma_5\gamma^i(-i\gamma^a\phi_a)
\gamma^j\gamma^b(i\partial_j\phi_b)\gamma^k\gamma^c(i\partial_k\phi_c)
+O(|x|^{-3})
\nonumber\\
&=\frac{1}{4\pi}\epsilon^{ijk}\epsilon^{abc}
\hat\phi_a\partial_j\hat\phi_b\partial_k\hat\phi_c +O(|x|^{-3}).
\nonumber
\end{alignat}
Substituting these into eq. (\ref{ModIndThe}), we obtain 
\begin{alignat}1
{\rm ind}\,{\cal H}=
\left\{
\begin{array}{ll}
\displaystyle{
-\frac{1}{2\pi}\int dx^i\epsilon^{ab}\hat\phi_a\partial_i\hat\phi_b
}
&\quad (d=2)\\
\displaystyle{
-\frac{1}{8\pi}\int dS_i
\epsilon^{ijk}\epsilon^{abc}\hat\phi_a\partial_j\hat\phi_b\partial_k\hat\phi_c
}
&
\quad (d=3)
\end{array}
\right. .
\label{WinNum}
\end{alignat}
Since in eq. (\ref{WinNum}), $\bm{\hat\phi}$ is a unit vector with $d$ components 
defined at $|x|\rightarrow\infty$ of R$^d$, $\bm{\hat\phi}$ 
can be regarded as a mapping from S$^{d-1}$ to S$^{d-1}$.  
It is well-known in mathematics that $\pi_{d-1}$(S$^{d-1}$)=Z, implying that
such a mapping is classified by integers. 
The r.h.s. of eq. (\ref{WinNum}) denotes exactly the winding number of 
$\bm{\hat\phi}$ over S$^{d-1}$ at $|x|\rightarrow\infty$.

In the case of the Fu-Kane model,\cite{FuKan08} 
we can choose $\bm{\hat\phi}=(\cos q\theta,\sin q\theta)$ at $r\rightarrow\infty$
if we consider a generic vortex with vorticity $q$. 
Substituting this into eq. (\ref{WinNum}), 
we then see that the winding number in eq. (\ref{WinNum}) is exactly $-q$, 
implying that the Hamiltonian has at least $q$ zero modes. 
It is not difficult to solve the eigenvalue equation and to show that
such a vortex indeed admits exactly $q$ zero modes. 
The r.h.s. of eq. (\ref{WinNum}) in the case of $d=3$ has been derived 
by Teo and Kane \cite{TeoKan09}
as a topological invariant for the many-body ground state of the model.
They have calculated the third Chern number \cite{QHZ08} associated with 
non-Abelian Berry's connection and shown that it reduces to the 
winding number of the order parameter over S$^2$. 
In the present calculations, without
resorting to adiabatic assumptions, we have established the index theorem
that is directly related to the analytic invariant associated 
with the Majorana zero modes with 
the topological invariant of the order parameter.

Thus far, we have derived the index theorem for a superconductor-topological insulator
model in the case where it has not only particle-hole symmetry but also chiral symmetry.
However, let us consider the term ${\cal H}'=i\gamma^a\gamma^b n_{ab}$,
for example. We see
that it transforms as $C{\cal H}'C^{-1}=-{\cal H}'$, but
$\gamma_5{\cal H}'\gamma_5=+{\cal H}'$. 
Therefore, generic Bogoliubov-de Gennes Hamiltonians have broken chiral symmetry.
In particular, the chemical potential is one of the most important terms 
breaking chiral symmetry when we study the superconductivity.
Hence, we will next investigate the effect of the chemical potential term
$-\mu\beta$ that breaks chiral symmetry, where $\beta$ is a matrix with 
the transformation properties $C\beta C^{-1}=-\beta$ and $\gamma_5\beta\gamma_5^{-1}=+\beta$.
\cite{footnote2}
When $\mu$ is a small constant, the perturbation theory indicates that the leading order 
corrections of the energies are given by the eigenvalues of the matrix
${\cal H}_{ij}'=-\mu\langle\varphi_{0,i}|\beta|\varphi_{0,j}\rangle$, where
$|\varphi_{0,i}\rangle$ denotes the ket notation of the unperturbed $i$th
zero mode wave function.
In general, we can choose $|\varphi_{0,j}\rangle=C|\varphi_{0,j}\rangle$, 
as already discussed. Then, we see that each matrix element is purely imaginary,
${\cal H}_{ij}'^{*}=-{\cal H}_{ij}'$ because of 
the particle-hole symmetry of $\beta$.
It thus turns out that ${\cal H}'$ is an antisymmetric 
purely imaginary matrix.
Without further symmetries, off-diagonal matrix elements of ${\cal H}'$ 
do no vanish generically.\cite{footnote3}  
This leads to an interesting suggestion \cite{TSL07} that
if an unperturbed ($\mu=0$) system admits several zero modes, some of them 
couple together through the chemical potential term
and become nonzero energy states such that two states form a pair 
with opposite energies. This is due to particle-hole symmetry.
Namely, an even number of zero modes become nonzero energy modes if the chemical potential
term is switched on.  
This implies that in a system with an odd number of unperturbed zero modes, 
there remains at least one zero mode even if any perturbations with particle-hole symmetry 
are added. 
In other words, the rank of an antisymmetric matrix is always even,
implying that odd-dimensional ${\cal H}'$ has inevitably an odd number of zero eigenvalues. 
We thus conclude that Majorana zero modes bound to a pointlike defect 
are classified by Z for class BDI, whereas they are classified by Z$_2$ for class D.
This is consistent with the results obtained by Tewari {\it et. al.}\cite{TSL07}

Finally, we present, as a nontrivial example,  
an explicit wave function of a zero mode in the Fu-Kane model\cite{FuKan08}
with a $q=1$ vortex. With the $\gamma$ matrices given in \cite{footnote1}, 
the zero mode wave function obeys
\begin{alignat}1
\left(
\begin{array}{cc}
i\sigma^j\partial_j-\mu&\phi_1-i\phi_2\\
\phi_1+i\phi_2&-i\sigma^j\partial_j+\mu
\end{array}
\right)
\left(\begin{array}{c}\xi\\ \eta\end{array}\right)=0 .
\nonumber
\end{alignat} 
Note that $\phi_1-i\phi_2=\Delta_0(r)e^{-i\theta}$, and that particle-hole symmetry 
makes it possible to choose $\eta=-\sigma_2\xi^*$.
\cite{footnote4}
Then, the above equation reduces to
\begin{alignat}1
\left(
i\sigma^j\partial_j-\mu
\right)\xi-\Delta_0(r)e^{-i\theta}\sigma^2\xi^*=0.
\label{FuKanZerMod}
\end{alignat}
If we further set $\xi^T=(v,u)$, we see that 
when $\mu=0$, two equations for $u$ and $v$ are decoupled, and 
the {\it normalizable} solution is given by $v=0$ and
\begin{alignat}1
u(r,\theta)\propto e^{-\int^r dr' \Delta_0(r')} ,
\nonumber
\end{alignat} 
which was obtained in Ref. \cite{FuKan08}.
Here, we have assumed that $\Delta_0(r)>0$ for $r\rightarrow \infty$.
This wave function has chirality $-1$, and hence, the index of this model is $-1$.
The above discussions on the classification scheme for classes BDI and D suggest that 
the present zero mode is stable against perturbations breaking chiral symmetry. 
Indeed, even when $\mu\ne0$, eq. (\ref{FuKanZerMod}) has
the following normalizable wave function:
\begin{alignat}1
&
\xi(r,\theta)\propto
e^{-\int^r dr'\Delta_0(r')}\left(
\begin{array}{c}
-ie^{-i\theta}J_1(\mu r)\\
J_0(\mu r)
\end{array}
\right) ,
\nonumber
\end{alignat}
where $J_n(x)$ is the Bessel function.

In conclusion, we have derived the index theorem  (\ref{WinNum}) for the model of a 
superconductor-topological insulator system in class BDI proposed by Kane {\it et al.}
This ensures the topological stability of the Majorana zero modes bound to
a topological defect.
For a model in class D, i.e., a model with particle-hole symmetry only, 
we have suggested that zero modes are classified by Z$_2$. Namely, 
only the parity (evenness or oddness) of the number of zero modes (the index, in other words)
in class BDI is invariant against chiral symmetry breaking perturbations.

We would like to thank A. Furusaki for useful comments on the universality
classes of topological insulators and superconductors, and also 
C. L. Kane and the Referee of JPSJ for pointing out the Z$_2$ classification of
zero modes in class D.
This work was supported in part by Grants-in-Aid for Scientific Research
(Nos. 20340098 and 21540378).


\begin{thebibliography}{99} 
%
\bibitem{ReaGre00}
N. Read and D. Green:
Phys. Rev. B {\bf 61} (2000) 10267.
%
\bibitem{Ivanov01}
D. A. Ivanov:
Phys. Rev. Lett. {\bf 86} (2001) 268.
%
\bibitem{SOM04}
A. Stern, F. von Oppen, and E. Mariani:
Phys. Rev. B {\bf 70} (2004) 205338.
%
\bibitem{SNT06}
S. Das Sarma, C. Nayak, and S. Tewari:
Phys. Rev. B {\bf 73} (2006) 220502.
%
\bibitem{TSL07}
S. Tewari, S. Das Sarma, and D.-H. Lee:
Phys. Rev. Lett. {\bf 99} (2007) 037001.
%
\bibitem{FuKan08}
L. Fu and C. L. Kane:
Phys. Rev. Lett. {\bf 100} (2008) 096407.
%
\bibitem{TYN09}
Y. Tanaka, T. Yokoyama, and N. Nagaosa:
Phys. Rev. Lett. {\bf 103} (2009) 107002.
%
\bibitem{STF09}
M. Sato, Y. Takahashi, and S. Fujimoto:
Phys. Rev. Lett. {\bf 103} (2009) 020401.
%
\bibitem{Vol09}
G. E. Volovik:
Pis'ma ZhETF {\bf 90} (2009) 639.
(arXiv:0909.3084)
%
\bibitem{Her09-1}
I. F. Herbut:
arXiv:0909.4231.
%
\bibitem{Her09-2}
I. F. Herbut:
arXiv:0910.4906.
%
\bibitem{TeoKan09}
J. C. Y. Teo and C. L. Kane:
arXiv:09094741.
%
\bibitem{QHZ08}
X.-L. Qi, T. L. Hughes, and S.-C. Zhang:
Phys. Rev. B {\bf 78} (2008) 195424. 
%
\bibitem{Kitaev03}
A. Kitaev:
Ann. Phys. {\bf 303} (2003) 2.
%
\bibitem{NSSFS08}
C. Nayak, S. H. Simon, A. Stern, M. Freedman, and S. Das Sarma:
Rev. Mod. Phys. {\bf 80} (2008) 1083. 
%
\bibitem{TSS09}
S. Tewari, J. D. Sau, and A. Das Sarma:
arXiv:0910.4763. 
%
\bibitem{footnote1}
In the Fu-Kane model, they have used
$\gamma^1=\sigma^1\otimes\sigma^3$,
$\gamma^2=\sigma^2\otimes\sigma^3$,
$\gamma^3=1\otimes\sigma^1$,
and $\gamma^3=1\otimes\sigma^2$, where in the notation of $a\otimes b$,
$a$ and $b$ denote the spin and particle-hole spaces, respectively.
The particle-hole transformation $C$ is defined as $C=\sigma^2\otimes\sigma^2 K$,
where $K$ denotes the complex conjugation operator.
In the Teo-Kane model,
$\gamma^j=\sigma^j\otimes\sigma^3\otimes\sigma^3$ $(j=1,2,3)$,
$\gamma^4=1\otimes\sigma^1\otimes1$,
$\gamma^5=1\otimes\sigma^2\otimes1$,
and $\gamma^6=1\otimes\sigma^3\otimes\sigma^1$,
where in $a\otimes b\otimes c$, $a$, $b$, and $c$ denote,
respectively, the spin, particle-hole, and orbital spaces.
For this model, $C=\sigma^2\otimes\sigma^2\otimes1 K$.
%
\bibitem{Zir96}
M. Zirnbauer:
J. Math. Phys. {\bf 37} (1996) 4986.
%
\bibitem{AltZir97}
A. Altland and M. Zirnbauer:
Phys. Rev. B {\bf 55} (1997) 1142.
%
\bibitem{SRFL08}
A. P. Schnyder, S. Ryu, A. Furusaki, and A. W. W. Ludwig:
Phys. Rev. B {\bf 78} (2008) 195125;
arXiv:0905.2029.
%
\bibitem{Kitaev08}
A. Kitaev:
arXiv:0901.2686.
%
\bibitem{JacRos81}
R. Jackiw and P. Rossi:
Nucl. Phys. {\bf 190} (1981) 681.
%
\bibitem{Wei81}
E. J. Weinberg:
Phys. Rev. D {\bf 24} (1981) 2669.
%
\bibitem{tHooft}
G. 't Hooft:
Nucl. Phys. B {\bf 79} (1974) 276.
%
\bibitem{Polyakov}
A. M. Polyakov:
JETP Lett. {\bf 20} (1974) 194.
%
\bibitem{Callias78}
C. Callias:
Commun. Math. Phys. {\bf 62} (1978) 213.
%
\bibitem{FujSuz05}
For a review, see, e.g., 
K. Fujikawa and H. Suzuki:
{\it Path Integrals and Quantum Anomalies} 
(Oxford University Press, Oxford, 2004).
%
\bibitem{footnote2}
In the Fu-Kane model, $\beta=-i\gamma^3\gamma^4$.
%
\bibitem{footnote3}
If the matrix $\beta$ has chiral symmetry, 
${\cal H}'$ vanishes in general.
This alternatively indicates the stability of
the index (\ref{WinNum}).
%
\bibitem{footnote4}
This condition is represented as $\varphi^c\equiv C\varphi=-i\varphi$,
where $\varphi^T=(\xi^T,\eta^T)$. The factor $-i$ is for notational convenience.

\end{thebibliography}
\end{document}